\documentclass[aps,prl,twocolumn,floatfix, groupedaddress]{revtex4}
\usepackage{graphicx}

\begin{document}

\title{Comment on
"Critical behavior of the Pauli spin susceptibility..." \\
by A.~A.~Shashkin \emph{et al}., cond-mat/0409100}
\author{M.\ Reznikov and U.\ Sivan}
\affiliation{Department of Physics and Solid State Institute,
Technion-IIT, Haifa 32000, Israel}

\begin{abstract}
The paper by A.~A.~Shashkin~\emph{et~al}. reports measurements of
the thermodynamic magnetization of two-dimensional electrons in
silicon. Although the experimental data is very similar to that
reported by us~\cite{Prus02} more then two years ago, the authors
arrive at an opposite conclusion regarding the spin susceptibility
"critical behavior" and spin instability in the vicinity of the
metal-insulator transition. We show that this interpretation is
based on a flawed analysis of the experimental data.

\end{abstract}

\maketitle

The state of strongly correlated fermions at low temperatures, and
in particular its magnetic properties, that has been a subject of
active interest, is still poorly understood. Some theories predict
that in the absence of disorder exchange effects can lead to
spontaneous magnetization, but such instability has never been
observed, see~\cite{Prus02} for references. Several years ago we
proposed a novel technique for measuring the thermodynamic
magnetization of 2d electrons and applied it to a high mobility
silicon MOSFET. Notwithstanding the enhanced spin susceptibility
observed in the experiment, the magnetization vanished smoothly as
the magnetic field was reduced, thus excluding the possibility of a
spin instability at zero field.

Ref.~\cite{shas04} repeats our measurements and reproduces the data,
extending it to 20\% lower carrier concentration. Albeit the
similarity in the data, the conclusion of  Ref.~\cite{shas04} is
opposite to ours. Based on extrapolation from strong fields authors
infer magnetic instability at zero field. In what follows we show
how this erroneous conclusion emerges from a wrong assumption.

\begin{figure}[tbp]
\includegraphics [width=8cm] {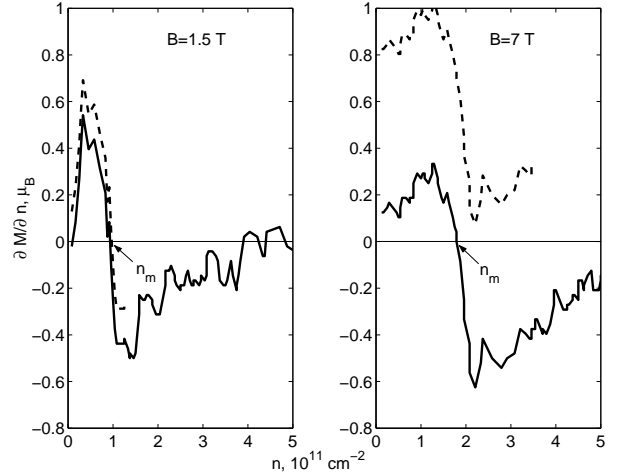}
\caption{Solid line: $\partial M/\partial n$ replotted
from~\cite{shas04}. Dashed line: the same data shifted by
$0.15\mu_B$ and $0.7\mu_B$ for $B=1.5$ and $7\,{\rm T}$ respectively
in order to account for the diamagnetic contribution of
$-0.1\mu_B/$~T. } \label{dmdn}
\end{figure}

The experimental method is described in Prus \emph{et
al.}~\cite{Prus02}. It yields directly $\partial \mu/\partial B$
which by the Maxwell relation equals $-\partial M/\partial n$, with
$\mu$, $n$ the electron chemical potential and density, and $M$, $B$
the magnetization and magnetic field. Fig.~2 in Ref.~\cite{shas04}
presents $\partial \mu/\partial B$ for $B$ between 1.5~T and 7~T.
The curves~\cite{shas04} for $B=1.5,\,7\,{\rm T}$ are reproduced as
$\partial M/\partial n$ by solid lines in Fig.~\ref{dmdn}. In
Ref.~\cite{shas04} the vanishing of $\partial \mu/\partial B$ is
identified with the full spin polarization. The corresponding $B$
values plotted versus density (Fig. 4a in Ref.~\cite{shas04}) align
along a straight line that, when extrapolated to $B=0$, crosses the
density axis at a finite $n$ close to the critical density $n_c(0)$
below which the system becomes strongly localized at $B=0$. From
such an extrapolation Ref.~\cite{shas04} concludes that the crossing
indicates spontaneous spin polarization at zero field.

\begin{figure}[tpb]
\includegraphics [width=8cm] {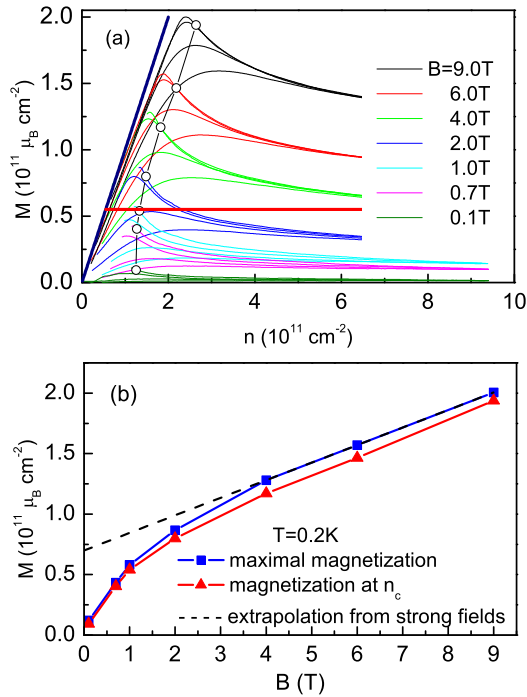}
\caption{(a) Spin magnetization as a function of density at
different magnetic fields and temperatures 0.2, 0.8, 2.5 and 4.2~K;
higher magnetization corresponds to lower temperature. Critical
densities, $n_c$, are marked by circles. Thick blue line - full
magnetization, thick red line - magnetization of a degenerate ideal
electron gas at $B=6$~T. (b) Maximal spin magnetization and spin
magnetization at the critical densities plotted against magnetic
field. Dashed line - extrapolation from high magnetic fields
(reproduced from~\cite{Prus02}).} \label{M(n)}
\end{figure}

We first note that the association of $\partial \mu/\partial B=0$
with full spin polarization is baseless. According to the Maxwell
relation, full spin polarization would occur at $\partial
\mu/\partial B=-\partial M/\partial n=-\mu_B$, where $\mu_B$ is the
Bohr magneton. In contrast, $\partial M/\partial n=0$ occurs at the
maximal magnetization rather then at the full one, and we denote the
corresponding density $n_m$. Fig.~\ref{M(n)}a, taken
from~\cite{Prus02}, depicts spin magnetization \emph{vs.} density
obtained by integration of $\partial M/\partial n$ data. We note
that the density $n_m(B)$ is always smaller than $n_c(B)$, and the
magnetization at this density, $M(B, n_m(B))$, is always below the
full one, $\mu_B n_m(B)$.

The degree to which the assumption of full spin polarization at
$n_m$ is invalid is also evident directly from the data of
Ref.~\cite{shas04}. The dependence $M(n_m)$ can be obtained  by
integration: $M(n_m)=\int\limits_{0}^{n_m} dn\partial M/\partial n
$. Since the measured $\partial M/\partial n$ (Fig.~\ref{dmdn})
never approaches $\mu_B$, the maximal magnetization is significantly
lower then the full one, and therefore the susceptibility determined
in~\cite{shas04} as $\mu_B n_m(B)/B$ and plotted in fig.~4b
of~\cite{shas04} is also significantly overestimated.

We also remark on the risk of drawing conclusions at $B=0$ based on
extrapolation from high $B$. It is quite easy to fall into the trap
of arguing for instability at low $B$ by attempting to judge on the
faith of the maximal magnetization or the magnetization at $n_c$ by
the behavior at high field. However, as clearly seen in
Fig.~\ref{M(n)}b, both go smoothly to zero with vanishing magnetic
field.

Finally we note that one should not ignore the diamagnetic
contribution to $M$~\cite{estimate}. This contribution can be
roughly estimated from the experimental data by assuming that the
maximal magnetization at high fields, e.g. 7~T, approaches $1\mu_B$
per electron. To obtain such value from Fig.~\ref{dmdn} the curve
should be shifted by approximately $0.7\mu_B$. This will also
increase $\partial M/\partial n$ for 1.5~T, keeping it still well
below the full spin polarization value (see dashed lines in
Fig.~\ref{dmdn}). This approximation for the diamagnetic shift is
supported by theoretical calculations~\cite{Prus02}.

In conclusion, the data presented in Ref.~\cite{shas04}, properly
analyzed, indicates that the 20\% reduction in the accessible
density compared to the earlier studied samples does not lead to
spin instability. Such an instability, anticipated by theory, may
still occur at a much lower density.


\begin{thebibliography}{qqq}
\bibitem{estimate}
For parabolic confining potential $U(x)=m^\ast\omega_0^2x^2/2$, in
the presence of parallel magnetic field with the cyclotron frequency
$\omega_c$, the electron energy is
$E_0=\frac12\hbar(\omega_0^2+\omega_c^2)^{1/2}$. Expanding
$(\omega_0^2+\omega_c^2)^{1/2}=\omega_0+\omega_c^2/(2\omega_0)$,
with $\partial \omega_c/\partial B=e/m^*c$, the ground state energy
field dependence is
%
\[
\partial E_0/\partial B\approx (m/m^*)\times (\omega_c/\omega_0)
\times\mu_B
\]
%
The diamagnetic contribution to the energy of spatial quantization
is linear in B, as expected. For the chemical potential set by $E_0$
we obtain $\partial \mu /\partial B \approx \partial E_0/\partial
B$. The ratio $\omega_c/\omega_0$ can be estimated as
$(d/\lambda)^2$, with $d$ the 2DEG thickness and $\lambda$ the
magnetic length. For $d\approx 30 {\AA}$ and $B=10$~T the
diamagnetic contribution is comparable to $\mu_B$ and thus cannot be
ignored.
\end{thebibliography}
\end{document}